\documentstyle[psfig]{l-aa}

\newcommand{\etal}{et al }
\thesaurus{11 (11.06.1; 11.17.2; 11.17.4: BRI~1335-0415)
         12 (12.03.3;12.05.1)
         13 (13.19.1) }

\title{The second detection of CO at redshift larger than 4}

\author{S. Guilloteau \inst{1}
\and A. Omont \inst{2}
\and R.G. McMahon \inst{3}
\and P. Cox \inst{4}
\and P. Petitjean \inst{2}
}

\institute{
IRAM, 300 rue de la Piscine, F-38406 Saint Martin d'H\`eres Cedex
\and
Institut d'Astrophysique de Paris, CNRS, 98bis Bd Arago, F-75014 Paris
\and
Institute of Astronomy, Madingley Road, Cambridge CB3 0HA, U.K.
\and
Institut d'Astrophysique Spatiale, Universit\'e Paris 11, F-91406 Orsay Cedex
}

\offprints{S.Guilloteau; email:guillote@iram.fr}
\date{Accepted for publication in Astronomy and Astrophysics}

\begin{document}

\sloppy

\maketitle

\begin{abstract}
  
  We report the detection with the IRAM interferometer at 3\,mm of
  J=5-4 CO line in the radio quiet quasar BRI~1335-0415 at redshift of
  $z=4.41$.  After BR~1202-0725 at $z=4.69$ (Ohta \etal 1996, Omont
  \etal 1996a), this is the second detection of CO at $z>3$. The
  integrated line intensity is $2.8 \pm 0.3$~Jy~km~s$^{-1}$ with a
  linewidth of 420$\pm$60~km~s$^{-1}$.  The dust continuum emission
  has also been mapped at 1.35~mm.  The 1.35~mm flux is found to be
  5.6$\pm$1.1~mJy.  The ratio of the CO to 1.35\,mm continuum flux is
  slightly larger than for BR~1202-0725.  Contrary to the case of
  BR~1202-0725, there is only marginal evidence of extension of the
  1.35\,mm continuum and 3mm CO emission.  In the absence of
  gravitational lensing, for which there is no a priori evidence, and
  within the uncertainties of the CO to M(H$_2$) conversion factors,
  the mass of molecular gas M(H$_2$) could be $10^{11}$M$_\odot$.

\end{abstract}

\section{Introduction}

The detection of CO and dust at millimeter wavelength in a few
prominent objects at very large redshift opens the possibility to
study the molecular gas of the primordial starbursts of galaxies. The
initial detection of CO in FIRAS10214+4724 at $z=2.3$ (Brown and
Vanden Bout 1992, Solomon \etal 1992), followed by CO detections in
H1413+117 (The Cloverleaf) at $z=2.5$ (Barvainis \etal 1994),
BR~1202-0725 at $z=4.7$ (Ohta \etal 1996, Omont \etal 1996a) and,
recently, in 53W002 at $z=2.34$ (Scoville \etal 1997).  In addition
thermal emission has now been convincingly detected at 1.3\,mm in
$\sim$8 $z>2$ QSOs (Barvainis \etal 1992; McMahon \etal 1994, Isaak
\etal 1994, Ivison, 1995, Omont et al 1996b) and in one radiogalaxy at
$z > 2$ (Dunlop \etal 1992, Chini \& Kr\"{u}gel 1994; Hughes \etal
1997).

Such a jump in the redshift range of millimeter radioastronomy was
made possible by~: i) the existence at larger redshifts of
extraordinary objects with molecular gas, dust and huge starbursts
similar to the most powerful ultra-luminous IRAS galaxies; ii) the
large increase of dust and CO luminosities with redshift when observed
at fixed wavelength; iii) the existence of a spectacular gravitational
lensing amplification in the two prominent cases of FIRAS10214+4724
and H1413+117.  However, since some tentative detections of quasars
and quasar absorption line systems have not been confirmed (Wiklind \&
Combes, 1994; Braine, Downes \& Guilloteau, 1996), the number of
detections of CO at $z>2$ remains small, consisting of the four
detections previously mentioned.  It is important to extend the number
of CO detections at large redshift so that we can study more
systematically: the nature of these sources, the properties of and the
relative amount of molecular gas and dust, the spatial distribution of
CO and dust, the role of gravitational amplification, and the spatial
extension of their millimeter emission (e.g. is the extension observed
in BR~1202-0725 unique ?)

In order to address these issues, we have adopted the following
strategy: i) search for dust emission toward radio-quiet quasars at
very high redshift in the millimeter or submillimeter ranges using
single dish telescopes; ii) for the sources detected, systematically
search for CO and map the continuum with the IRAM
interferometer. Among the detections reported by Omont \etal (1996b),
the most obvious candidate to search for CO with the interferometer
was BRI~1335-0415, an optically selected $z=4.4$ QSO from the APM
multicolour quasar survey (Irwin \etal 1991, Storrie-Lombardi \etal
1996). It is the second source with $z>4$ with the strongest 1.3mm
dust emission after BR~1202-0725 (Omont \etal 1996b). We report in
this letter the detection of CO(5-4) in BRI~1335-0415 with an
intensity comparable to that of BR~1202-0725.

\section{The redshift of BRI~1335-0415}

Because of the relatively small instantaneous bandwidth of available
heterodyne receivers (500\,MHz at 105\,GHz, i.e. 1400 km~s$^{-1}$ or
$\Delta z \sim0.02$ at $z=4.5$, a prerequisite for CO observations is
a precise determination of the the redshift of the target object.

In the case of high redshift quasars the redshifts have to be
determined from rest frame ultra-violet emission lines (e.g.
Ly-$\alpha$:1216\AA, C{\sc v}:1549\AA) with widths of $\sim$10,000
km~s$^{-1}$.  Moreover, it is well established that the above high
ionization lines show systematic differences when compared with
redshifts derived from lower ionization lines (Carswell \etal 1991,
Tytler \& Fan 1992).  Espey \etal (1990) give a mean redshift
difference of 1000~km~s$^{-1}$ in $z<3$ quasars.  Of particular
relevance to this work, Storrie-Lombardi \etal (1996) find
430$\pm$60~km~s$^{-1}$ for the APM sample of $z>4$ quasars.  The error
quoted above is the uncertainty in the mean and the actual
dispersion(1$\sigma$) about the above value is 1260~km~s$^{-1}$.

We have made an independent redetermination of the redshift of
BRI~1335$-$0415 using the same data as Storrie-Lombardi \etal (1996).
The C~{\sc iv} line gives $z=4.370$ and the Si~{\sc iv} line
$z=4.378$, while the low ionisation lines O~{\sc i} and C~{\sc ii}
give $z=4.410$ and 4.426, respectively.  However, the O~{\sc i} and
C~{\sc ii} measurements are certainly not as reliable as the Si~{\sc
iv} and C~{\sc iv} ones, as the former lines are quite
weak. Therefore, from the emission lines, one can just estimate that
the redshift should lie in the broad range 4.370--4.426.

However, as in the case of BR~1202-0725 (Isaak \etal 1994) another
indication can be derived from the absorption lines at redshifts close
to that of the quasar. On the line of sight of BRI~1335-0415, there is
a Lyman limit system (LLS) quite close to the quasar redshift
(Storrie-Lombardi \etal 1994, 1996) with an estimated H~{\sc i} column
density of the order of 10$^{18}$~cm$^{-2}$ Its redshift can be
established from the presence of the C~{\sc ii} $\lambda 1334$ and
Si~{\sc ii} $\lambda 1260$ lines detected at $z=4.403$ and 4.406,
respectively. There is no doubt that these lines are due to the LLS.
In addition, there is a strong Ly$\alpha$ line at $\lambda \sim
6580$~\AA, i.e. $z \sim 4.41$. These facts confirm that the LLS is at
$z \sim 4.405$.  This value can appear as one of the best guess for an
approximate value of the redshift of the quasar. Lyman-limit
absorption systems so close to the redshift of the quasar are rare and
only occur within 4000 km~s$^{-1}$ in 2 of 15 quasar sight lines
studied by Storrie-Lombardi \etal (1996).

\section{Observations}
Observations were made with the IRAM interferometer between May 1996
and March 1997. We used the standard CD configuration, which gives a
beam of $2.6^{\prime\prime} \times 1.8^{\prime\prime}$ at PA
24$^\circ$ at 1.35\,mm and $6.1^{\prime\prime} \times
3.4^{\prime\prime}$ at 3\,mm.  Dual frequency receivers were used to
search simultaneously for CO(5-4) emission at 3mm and dust emission at
1.35\,mm.  Two different tunings, separated by 450 MHz, were used at
3\,mm, at 107.020 GHz and 106.570 GHz to provide wide velocity
coverage. The 3\,mm receivers were tuned in single sideband and the
1.35\,mm receivers in double sideband (at 223 GHz LSB and 226 GHz
USB). Typical SSB system temperature were $\simeq 150$~K at 3\,mm and
$\simeq 350$~K at 1.35\,mm.  Amplitude and phase calibration were done
using 3C273, whose flux raised from 19 to 23 Jy at 106 GHz, and 15 to
19 Jy at 224 GHz during that period. Flux density scale is accurate to
better than 10\%.  Phase noise was below 30$^\circ$ on all baselines
even at 1.35\,mm.  The final results are presented in Figs.1-3.  The
total integration time is about 10 hours for the CO(5--4) line, and 18
hours for the 1.35\,mm continuum.

\begin{figure}
\psfig{file=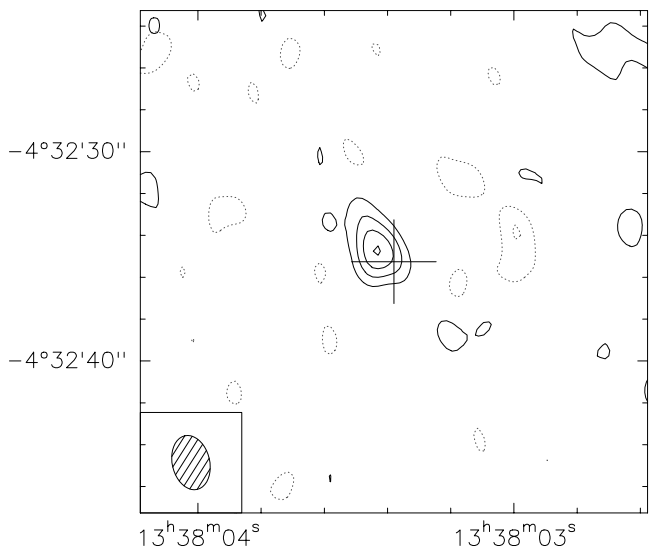,width=8.5cm,angle=270.0}
\caption[]{1.35\,mm continuum map of BRI1335-0415. Contour step is 1 mJy/beam,
and the rms noise 0.7 mJy/beam. The cross indicates the optical position.}
\label{fig:1}
\end{figure}

\begin{figure}
\psfig{file=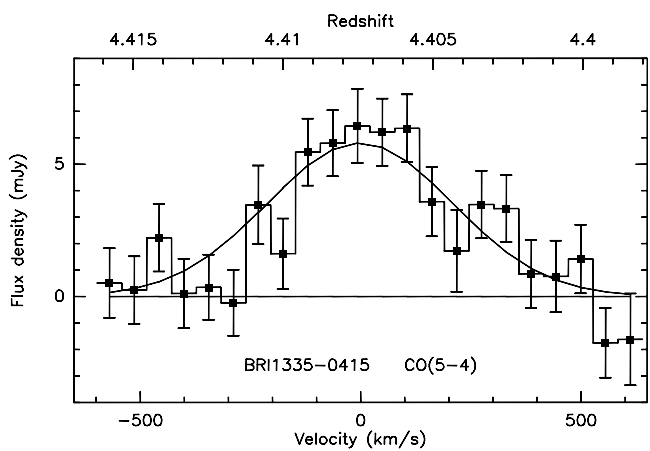,width=8.5cm,angle=270}
\caption[]{Spectrum of the CO J=5-4 line towards BRI~1335-0415, superimposed
with the best Gaussian profile. Errorbars are $\pm 1 \sigma$.
The velocity scale corresponds to a frequency of 106.570 GHz corresponding
to a redshift $z=4.4074\pm0.0015$.}
\label{fig:3}
\end{figure}

\begin{figure*}
\psfig{file=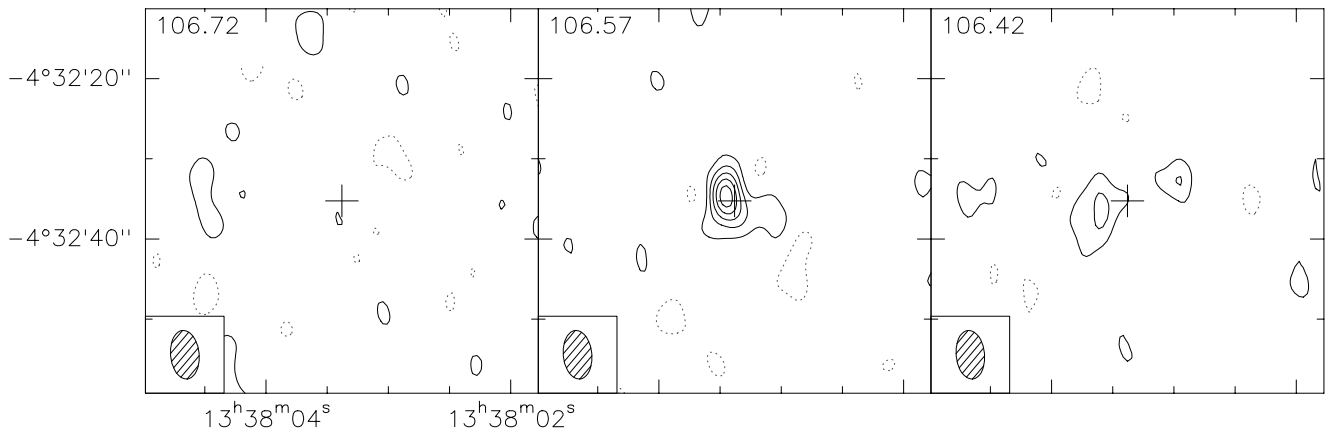,width=17.5cm,angle=270.0}
\caption[]{Channel maps of the CO J=5-4 line towards BRI~1335-0415.
Channel width is 150 MHz, i.e. 420 km~s$^{-1}$. The contour step is
0.8 mJy/beam (2 $\sigma$). The observing frequency in GHz is indicated in the
upper left corner of each map.}
\label{fig:2}
\end{figure*}




At 1.35 mm, we detected a total flux of $5.6 \pm 1.1$~mJy (Fig.\,1).
The continuum source position is offset by ($0.7 \pm
0.2^{\prime\prime},0.5 \pm 0.2^{\prime\prime}$) from the optical
position (RA $13^h38^m03.38^s$ Dec
$-04^\circ32^{\prime}35.3^{\prime\prime}$ J2000.0), well within the
astrometric uncertainty of both measurements.  The source appears only
marginally resolved, with a formal gaussian size of $1.0 \pm
0.4^{\prime\prime}$. This could be purely a seeing effect.

While the setting at 107.020 GHz only yields an upper limit of 1.2 mJy
(3 $\sigma$) on the 3\,mm continuum, the CO(5-4) line
($\nu_{rest}$=576.2677~GHz) is clearly detected towards the position
of the continuum emission.  The CO(5--4) line is found at a frequency
of 106.570~GHz, corresponding to a redshift of $4.4074 \pm
0.0015$. The spectrum is displayed in Fig.~2, and maps over 150 MHz
wide channels are shown in Fig.~3.  The integrated line flux is $2.8
\pm 0.3$~Jy~km~s$^{-1}$ with a line width of $420 \pm 60$~km~s$^{-1}$.

The broad channel maps of the CO(5--4) line (Fig.~3) show marginal
evidence for extension and/or position shift with velocity. The first
contour (2~$\sigma$ levels) is clearly extended in the central
channel centered at 106.57 GHz, whereas the 2~$\sigma$ contour is seen
towards the east in the channel at 106.42 GHz. In both channels, the
emission extends about one beam away from the peak emission. However,
it is difficult to decide with the present signal to noise ratio
whether this result is really significant.

\section{Discussion.}
There is no doubt about the reality of the CO detection in
BRI~1335-0415: the signal-to-noise ratio is high, especially on the
integrated intensity. The profile, as well as the baseline, is very
well behaved. The coincidence with the QSO position and the absence of
signal elsewhere (Fig.\ref{fig:2}) further strengthens the reality of
the detection.  The lack of pure continuum emission when all the 3-mm
broad band measurements, except in the line frequency range, are
included, indicates that the detected 3mm emission comes from a
spectral line (Fig.\ref{fig:3}).  The width of 420 km~s$^{-1}$ is in
the range of, although slightly larger than, the values observed for
other high--redshift sources~: 200-350 km~s$^{-1}$ for BR~1202-0725
(Omont \etal 1996a), 220 km~s$^{-1}$ in F10214+4724 (Solomon \etal
1992), 330 km~s$^{-1}$ in H1413+117 (Barvainis \etal 1994), from 150
to 550 km~s$^{-1}$ for a sample of 37 ultraluminous galaxies (Solomon
\etal 1997).

The derived redshift for the molecular gas is just in the redshift
range derived from optical emission lines, in the upper part as
expected, close to the low ionisation lines. Indeed, it is practically
coincident with the redshift of the Lyman limit absorption system
(LLS).  It will be important to obtain a better redshift for this
Lyman limit system to see if the redshift agrees more closely with the
CO.

The integrated CO intensity of BRI~1335-0415 (2.8 Jy km~s$^{-1}$) is
slightly stronger than for BR~1202-0725 (2.4 Jy km~s$^{-1}$) which is
at a similar redshift.  As in other sources (see, e.g., Omont \etal
1996a), several approximations are required to estimate the mass of
molecular hydrogen in BRI~1335-0415 from the CO intensity. Assuming a
relatively high rotational temperature demonstrated in other sources
by the CO line ratios, and using the same conversion factor as in
Omont \etal (1996a) for the mass of molecular hydrogen from the (5-4)
line luminosity (see Solomon \etal 1992), we obtain $M({\rm H}_2) = \,
6 \times 10^{10}h^{-2}{\rm M}_{\odot}$ where h~=~H$_{\rm o}$/100 km
s$^{-1}$ Mpc$^{-1}$ and adopting q$_{\rm o}$~=~0.5 . This is similar
to the value derived for BR~1202-0725 in Omont \etal (1996a). However,
this H$_2$ mass could be overestimated as in the case of ultraluminous
IR galaxies with comparable CO luminosities where the H$_2$ masses,
derived as above, are too high by a factor 3 (Solomon \etal 1997,
Downes \etal 1993).

The measured total continuum flux at 1.35\,mm ($5.6 \pm 1.1$~mJy), is
consistent with the value measured at an effective wavelength of
$\sim$1.25\,mm at the IRAM 30--m with the bolometer receivers ($10.3
\pm 1.35$~mJy), when the high spectral index (3.5--4) of the thermal
dust emission is taken into account.  However, it is strange that the
ratio of the 1.35\,mm fluxes of BR~1202-0725 and BRI~1335-0415 as
measured with the interferometer (Omont \etal 1996a and this paper) is
2.8$\pm$0.4, while the ratio of the 1.25\,mm fluxes measured with the
30--m telescope is 1.2$\pm$0.2 (Omont \etal 1996b).  It is important
to understand why there is such a difference. Differences in the dust
temperature or a range in the dust to molecular gas ratio are two
possibilites that can be addressed by further observations. Finally,
although the present 30--m and interferometer values for the continuum
flux of BRI~1335-0415 are consistent, the smaller flux measured with
the interferometer could indicate that the interferometric
observations miss a weak extended component of the emission.

In several other aspects, BRI~1335-0415 appears notably different from
other high $z$ sources, and especially from BR~1202-0725. First there
is no evidence for a large extension of the 1.35~mm continuum emission
with respect to the 2$''$ beam (Fig.~1), while in BR~1202-0725 there
is another strong component 4$''$ away from the QSO. A priori, in
BRI~1335$-$0415 there is no indication of gravitational lensing as is
clearly the case for F10214+4724 and H1413+117. Second, the line of
sight of BRI~1335-0415 also appears less rich in absorption systems
than BR~1202-0725.  Storrie-Lombardi \etal (1996) report $\sim$10
unique intervening redshift systems in BR~1202-0725 whereas in the
case of the BRI~1335$-$0415 they report only 2 such systems, i.e. the
$z=4.404$ Lyman-limit system discussed above and a strong Mg~{\sc
ii}/Fe~{\sc ii} system at $z = 1.823$.

\section{Conclusion}
With a CO intensity comparable to that of BR~1202-0725, BRI~1335-0415
is another exceptional object at redshift larger than 4.  If the
absence of strong gravitational magnification is firmly established,
it could have one of the largest mass of molecular gas known,
$\sim10^{11}$M$_\odot$. Such a large mass of molecular gas probably
indicates a huge starburst, possibly associated to the formation of
the core of an elliptical galaxy or bulge of a massive spiral
galaxy. The coincidence of the redshifts shows that the CO emission
and the Lyman limit absorption system may pertain to the same object,
possibly the host galaxy of the QSO. If so a more detailed chemical
analysis of the absorption line system would be very interesting since
it would allow a determination of the metallicity of the interstellar
medium in this quasar's host galaxy.

Contrary to BR~1202-0725, neither the 1.35~mm emission, nor the CO
emission display a strong extension at a scale of several
arcseconds. However, there is an indication of a weak extension of the
CO emission which should be confirmed. It would also be interesting to
further compare the continuum 1.3\,mm continuum flux measured with the
IRAM 30--m telescope and interferometer. Although the present values
are consistent, it is not excluded that the smaller flux measured with
the interferometer could indicate that it misses a weak extended
component of the emission. It will also be important to understand why
the ratio of the CO intensity to the 1.35~mm interferometer flux is
three times larger in BRI~1335-0715 than in BR~1202-0725.

\begin{acknowledgements}
This work was carried out in the context of EARA, a European Association 
for Research in Astronomy. RGM thanks the Royal Society for support. We are 
grateful to the IRAM staff at Bure for its efficient assistance.
\end{acknowledgements}


\end{document}